\documentclass[10pt,twoside,english,aps,prl,superscriptaddress,twocolumn,notitlepage,10pt]{revtex4-1}
\usepackage{mathpazo}

\usepackage[LGR,T1]{fontenc}
\usepackage[latin9]{inputenc}
\usepackage[a4paper]{geometry}
\usepackage{xcolor}
\usepackage{pdfcolmk}
\usepackage{babel}
\usepackage{verbatim}
\usepackage{textcomp}
\usepackage{graphicx}
\PassOptionsToPackage{normalem}{ulem}
\usepackage{ulem}
\usepackage[unicode=true,pdfusetitle,
 bookmarks=false,
 breaklinks=true,pdfborder={0 0 0},pdfborderstyle={},backref=false,colorlinks=true]
 {hyperref}
\hypersetup{
 colorlinks=true,citecolor=blue,linkcolor=blue,urlcolor=blue}

\makeatletter

\DeclareRobustCommand{\greektext}{%
  \fontencoding{LGR}\selectfont\def\encodingdefault{LGR}}
\DeclareRobustCommand{\textgreek}[1]{\leavevmode{\greektext #1}}
\ProvideTextCommand{\~}{LGR}[1]{\char126#1}

\providecolor{lyxadded}{rgb}{0,0,1}
\providecolor{lyxdeleted}{rgb}{1,0,0}

\DeclareRobustCommand{\lyxsout}[1]{\ifx\\#1\else\sout{#1}\fi}

\usepackage{soul}

\usepackage{graphicx}
\usepackage[calcwidth]{titlesec}

\makeatletter
\renewcommand\frontmatter@abstractwidth{\dimexpr\textwidth-2.8cm\relax}
\makeatother

\addto\captionsenglish{}

\titleformat{\section}[hang]{\large\bfseries}{}{0em}{}
\titlespacing\section{0pt}{4pt plus 2pt minus 2pt}{0pt plus 2pt minus 2pt}

\AtBeginDocument{
\renewcommand{\ref}[1]{\autoref{#1}}

}

\setcitestyle{super}

\makeatother

\begin{document}

\title{Tunable Room Temperature Magnetic Skyrmions in Ir/Fe/Co/Pt Multilayers\smallskip{}
}

\author{Anjan Soumyanarayanan}
\email{souma@dsi.a-star.edu.sg}

\affiliation{Data Storage Institute, 2 Fusionopolis Way, 138634 Singapore}

\affiliation{Division of Physics and Applied Physics, School of Physical and Mathematical
Sciences, Nanyang Technological University, 637371 Singapore}

\author{M. Raju}

\affiliation{Division of Physics and Applied Physics, School of Physical and Mathematical
Sciences, Nanyang Technological University, 637371 Singapore}

\author{A. L. Gonzalez Oyarce}

\affiliation{Data Storage Institute, 2 Fusionopolis Way, 138634 Singapore}

\author{Anthony K. C. Tan}

\affiliation{Division of Physics and Applied Physics, School of Physical and Mathematical
Sciences, Nanyang Technological University, 637371 Singapore}

\author{Mi-Young Im}

\affiliation{Center for X-ray Optics, Lawrence Berkeley National Laboratory, Berkeley,
California 94720, USA}

\affiliation{Department of Emerging Materials Science, DGIST, Daegu 42988, Korea}

\author{A. P. Petrovi\'{c}}

\affiliation{Division of Physics and Applied Physics, School of Physical and Mathematical
Sciences, Nanyang Technological University, 637371 Singapore}

\author{Pin Ho}

\affiliation{Data Storage Institute, 2 Fusionopolis Way, 138634 Singapore}

\author{K. H. Khoo}

\affiliation{Institute of High Performance Computing, 1 Fusionopolis Way, 138632,
Singapore}

\author{M. Tran}

\affiliation{Data Storage Institute, 2 Fusionopolis Way, 138634 Singapore}

\author{C. K. Gan}

\affiliation{Institute of High Performance Computing, 1 Fusionopolis Way, 138632,
Singapore}

\author{F. Ernult}

\affiliation{Data Storage Institute, 2 Fusionopolis Way, 138634 Singapore}

\author{C. Panagopoulos}
\email{christos@ntu.edu.sg}

\affiliation{Division of Physics and Applied Physics, School of Physical and Mathematical
Sciences, Nanyang Technological University, 637371 Singapore}
\begin{abstract}
\textbf{Magnetic skyrmions are nanoscale topological spin structures
offering great promise for next-generation information storage technologies.
The recent discovery of sub-100 nm room temperature (RT) skyrmions
in several multilayer films has triggered vigorous efforts to modulate
their physical properties for their use in devices. Here we present
a tunable RT skyrmion platform based on multilayer stacks of Ir/Fe/Co/Pt,
which we study using X-ray microscopy, magnetic force microscopy and
Hall transport techniques. By varying the ferromagnetic layer composition,
we can tailor the magnetic interactions governing skyrmion properties,
thereby tuning their thermodynamic stability parameter by an order
of magnitude. The skyrmions exhibit a smooth crossover between isolated
(metastable) and disordered lattice configurations across samples,
while their size and density can be tuned by factors of 2 and 10 respectively.
We thus establish a platform for investigating functional sub-50 nm
RT skyrmions, pointing towards the development of skyrmion-based memory
devices.}
\end{abstract}
\maketitle
\noindent \noindent \phantomsection \addcontentsline{toc}{section}{Introduction}

\noindent %
In conventional ferromagnets (FMs), the exchange interaction aligns
spins, and the anisotropy determines energetically preferred orientations.
Meanwhile, the Dzyaloshinskii-Moriya interaction\citep{Dzyaloshinsky1958a,Moriya1960}
(DMI), generated by strong spin-orbit coupling (SOC) and broken inversion
symmetry, induces a relative tilt between neighbouring spins. Magnetic
skyrmions \textendash{} finite-size two-dimensional (2D) 'whirls'
of electron spin \textendash{} form due to the competition between
these `winding\textquoteright{} DMI and `aligning\textquoteright{}
exchange interactions\citep{Bogdanov1994,Rossler2006,Muhlbauer2009,Yu2010a,Nagaosa2013}.
Skyrmions have several compelling attributes as prototype memory elements,
namely their: (1) nontrivial spin topology, protecting them from disorder
and thermal fluctuations\citep{Fert2013,Sampaio2013,Hagemeister2015},
(2) small size and self-organization into dense lattices\citep{Muhlbauer2009,Yu2010a,Heinze2011,Romming2013},
and (3) solitonic nature, enabling particle-like dynamics, manipulation
and addressability\citep{Schulz2012,Romming2013,Hagemeister2015}.
Originally magnetic skyrmions were discovered in non-centrosymmetric
compounds hosting bulk DMI\citep{Rossler2006,Muhlbauer2009,Yu2010a,Nagaosa2013}.
However, their emergence in thin multilayer films with interfacial
DMI is particularly exciting\citep{Fert1990,Bode2007} due to the
inherent tunability of magnetic interactions in 2D, and the material
compatibility with existing spintronic technology\textbf{\citep{Fert2013,Soumyanarayanan2016}. }

The DMI generated at interfaces between ultrathin FM layers and strong
SOC metals\citep{Bode2007,Heide2008,Cho2015a,Dupe2014,Yang2015a}
can host Néel-textured skyrmions \textendash{} first observed in epitaxial
monolayers of Fe on Ir(111)\citep{Heinze2011,Romming2013}. Whereas
such ultrathin films can stabilize small skyrmions ($\sim$8~nm)
only at low temperatures ($<30$~K), analogous multilayer films have
recently been shown to host $\sim$50-100~nm RT skyrmions\citep{Moreau-Luchaire2015a,Woo2015,Boulle2016}.
Here, a FM layer (e.g. Co) is sandwiched between different SOC metals
(e.g. Pt and Ir/Ta) to produce a net effective DMI\citep{Moreau-Luchaire2015a,Woo2015,Boulle2016},
and multiple repeats of such trilayers stabilize columnar skyrmions
through interlayer exchange coupling\citep{Chen2015,Nandy2016}. Multilayer
skyrmions have been imaged using synchrotron-based microscopy techniques\citep{Moreau-Luchaire2015a,Woo2015,Boulle2016},
with recent demonstrations of their confinement\citep{Moreau-Luchaire2015a,Boulle2016},
nucleation and dynamics\citep{Jiang2015,Woo2015,Buttner2015} in constricted
geometries. To translate these attributes into functional RT devices,
we must first develop methods of controlling and varying the physical
properties of skyrmions, such as their thermodynamic stability, size
and density. 

Individual skyrmion addressability will be particularly important
for technological applications. To realise this, we must achieve control
over the skyrmion stability via the critical material parameter, $\kappa$\citep{Bogdanov1994,Heide2008,Kiselev2011,Rohart2013,Leonov2016a}:
\begin{equation}
\kappa=\pi D/4\sqrt{AK}\equiv D/D_{{\rm c}}\ ,\label{eq:kappa}
\end{equation}
where $A$ is exchange stiffness, $D$ is the normalized DMI per unit
area, and $K$ is the out-of-plane anisotropy. For $\kappa>1$, skyrmions
are thermodynamically stable entities, forming a lattice at equilibrium,
while for $0<\kappa<1$, they would be metastable, isolated particles\citep{Kiselev2011,Rohart2013,Leonov2016a}.
It is also important to establish a route towards reducing the skyrmion
size at RT to maximize its potential for energy-efficient, high-density
memory. Furthermore, the ability to control skyrmion density would
enable device performance tuning in both static\citep{Sampaio2013}
and dynamic\citep{Tomasello2014,Zhang2015} configurations. Investigating
the technological relevance of multilayer skyrmions will also necessitate
their electrical detection\citep{Neubauer2009,Raicevic2011,Porter2014}
and imaging\citep{Milde2013} within stack configurations translatable
to practical devices. 

Here we describe our development of a new material platform \textendash{}
multilayer stacks of Ir/Fe/Co/Pt \textendash{} as a host of sub-50~nm
skyrmions with continuously tunable properties. By harnessing the
large and opposite signs of DMI generated from Fe/Ir\citep{Heinze2011}
and Co/Pt\citep{Yang2015a} interfaces, we achieve substantial control
over the effective DMI governing skyrmion properties. We first confirm
the presence of nanoscale RT skyrmions via established X-ray microscopy
techniques, and investigate their field evolution using magnetic force
microscopy (MFM) and Hall transport. By varying FM layer composition
and thickness, we control $D$, $K$ and $A$, and thereby modulate
the skyrmion stability parameter $\kappa$ by an order of magnitude,
the skyrmion size by a factor of 2, and the skyrmion density by a
factor of 10. We thus establish a platform for realizing tunable,
functional RT skyrmions in multilayers, and demonstrate how they may
be studied using commonly available lab-based techniques.

\noindent 
\section{Multilayer Stack Structure}

\noindent \textsf{\textbf{}}%

\begin{figure}[h]
\begin{centering}
\includegraphics[width=3.2in]{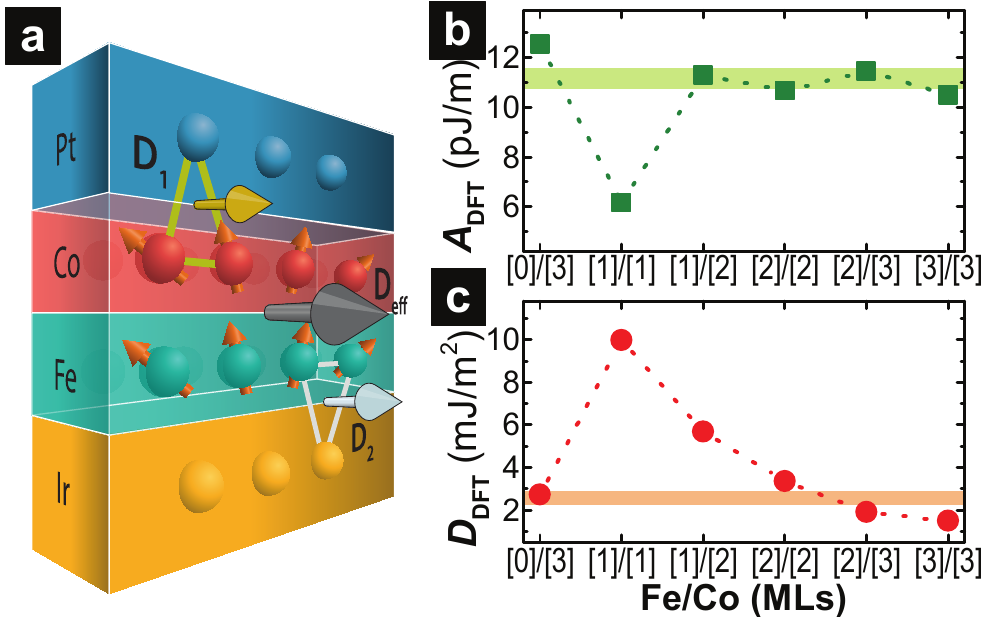}
\par\end{centering}
\noindent \caption{\textbf{DMI Enhancement in Ir/Fe/Co/Pt Stacks. (a) }Schematic of our
multilayer stack, featuring a sequence of Ir, Fe, Co, and Pt layers.\textbf{
}The large DMI vectors at Co/Pt (top, $\vec{D}{}_{{\rm 1}}$) and
Fe/Ir (bottom, $\vec{D}{}_{{\rm 2}}$) interfaces can act in concert
to enhance the effective DMI, $\vec{D}{}_{{\rm eff}}$.\textbf{ (b-c)
}DFT calculations of the exchange stiffness ($A_{{\rm DFT}}$, b)
and DMI ($D_{{\rm DFT}}$, c) for Ir{[}3{]}/Fe{[}$a${]}/Co{[}$b${]}/Pt{[}3{]}
stacks with varying Fe/Co composition (number of atomic layers in
braces). The light green line in (b) indicates the typical magnitude
of $A_{{\rm DFT}}$ for experimentally studied compositions. The addition
of Fe leads to an enhancement of $D_{{\rm DFT}}$ relative to Fe{[}0{]}/Co{[}3{]},
as shown by the light red line in (c).\textbf{ }\label{fig:Stack-DFT}}
\end{figure}

\noindent %
The interfacial DMI is defined for neighbouring spins $\vec{S}_{1,2}$
as 
\begin{equation}
\mathcal{H}_{{\rm DMI}}=-\vec{D}_{12}\cdot(\vec{S}_{1}\times\vec{S}_{2})\label{eq:DMI}
\end{equation}

\noindent The magnitude and sign of $\vec{D}_{12}$, determined by
the FM and SOC layers at the interface, govern the phenomenology of
multilayer skyrmions\citep{Fert2013,Nagaosa2013}. Notably, the Co/Pt
interface hosts a \textbf{\emph{large, positive DMI}} ($d_{{\rm Co-Pt}}^{{\rm tot}}\sim+3$~meV)\citep{Yang2015a,Boulle2016},
enabling Co-based trilayers with large DMI (Co/Pt) and small DMI (e.g.
Co/Ir\citep{Moreau-Luchaire2015a}, Co/MgO\citep{Boulle2016}, Co/Ta\citep{Woo2015}
etc.) interfaces to host skyrmions. Meanwhile, the Fe/Ir interface
hosts a \textbf{\emph{large, negative DMI}} ($d_{{\rm Fe-Ir}}^{{\rm tot}}\sim-2$~meV)\citep{Heinze2011,Dupe2014,Dupe2016,Yang2015a}.
Therefore, a stack structure combining the Co/Pt and Fe/Ir interfaces,
with large DMI of opposite signs, could exhibit additive enhancement
of the effective DMI. Here we examine such a four-layer Ir/Fe/Co/Pt
stack (\ref{fig:Stack-DFT}a) and establish it as a platform for tailoring
magnetic interactions and skyrmion properties. 

{} To validate our hypothesis of DMI enhancement, we performed \emph{ab
initio} density functional theory (DFT) calculations of the magnetic
interactions for the Ir/Fe/Co/Pt stack with varying Fe/Co composition.
We determined the effective DMI, $D_{{\rm DFT}}$, from the difference
between the DFT calculated energies for clockwise and counter-clockwise
chiral spin configurations, following the work of Yang \emph{et al.}\citep{Yang2015a,Boulle2016}
(details in Methods, \textcolor{blue}{§S2}). $D_{{\rm DFT}}$, shown
in \ref{fig:Stack-DFT}c, has a prominent peak for Fe{[}1{]}/Co{[}1{]}
(number of atomic layers in braces), supporting the hypothesis of
DMI enhancement with the incorporation of an Fe layer. Importantly,
for the same total FM thickness (e.g. Fe{[}1{]}/Co{[}2{]} c.f. Fe{[}0{]}/Co{[}3{]}),
we obtain a substantial ($\sim100\%$) DMI enhancement with Fe, which
persists even for larger FM thicknesses (e.g. Fe{[}2{]}/Co{[}2{]}).
During the preparation of this manuscript, we became aware of a similar
result reported for one such composition (Fe{[}1{]}/Co{[}2{]})\citep{Yang2016}.
Meanwhile, the exchange stiffness, $A_{{\rm DFT}}$ (\ref{fig:Stack-DFT}b),
falls in a narrow range between 10.5-12.5~pJ/m for Fe/Co compositions
corresponding to the experimental work. A notable exception to the
trend is Fe{[}1{]}/Co{[}1{]} ($A_{{\rm DFT}}\simeq6.2$~pJ/m) \textendash{}
such lowering of $A_{{\rm DFT}}$ for magnetic monolayers is well
documented\citep{Heinze2011,Dupe2014,Dupe2016}. Finally, increasing
the Fe/Co ratio and the FM thickness would also result in reduced
anisotropy\citep{Johnson1999}. Thus, varying the Fe/Co composition
could enable the modulation of $D$, $K$, and $A$ \textendash{}
thereby tailoring the ensuing skyrmion properties. 

Multilayer films with {[}Ir(10)/Fe($x$)/Co($y$)/Pt(10){]}$_{20}$
stacks (layer thickness in Å in parentheses) were sputtered on thermally
oxidized Si wafers with optimized film texture and interface quality,
and on Si$_{3}$N$_{4}$ membranes for X-ray microscopy (see Methods).
The thickness of Fe ($x:$ 0-6~Å) and Co ($y:$ 4-6~Å) layers were
varied across the films, and the samples studied are described in
terms of their \textbf{Fe($x$)/Co($y$)} composition. While the results
detailed here were measured in films with 20 stack repeats for enhanced
X-ray contrast, similar trends were observed for transport and MFM
in samples with 8 repeats. The data presented here were acquired following
saturation at positive fields ($H>\left|H_{{\rm S}}\right|$) in out-of-plane
(OP) configuration. The OP magnetization ($M(H)$) loops for these
stacks have a characteristic sheared shape (see e.g. \ref{fig:MTXM-MFM}j),
which is typically a signature of labyrinth domain states\citep{Woo2015}.

\section{Microscopic Imaging of RT Skyrmions}

\begin{figure*}[t]
\begin{centering}
\includegraphics[width=6.6in]{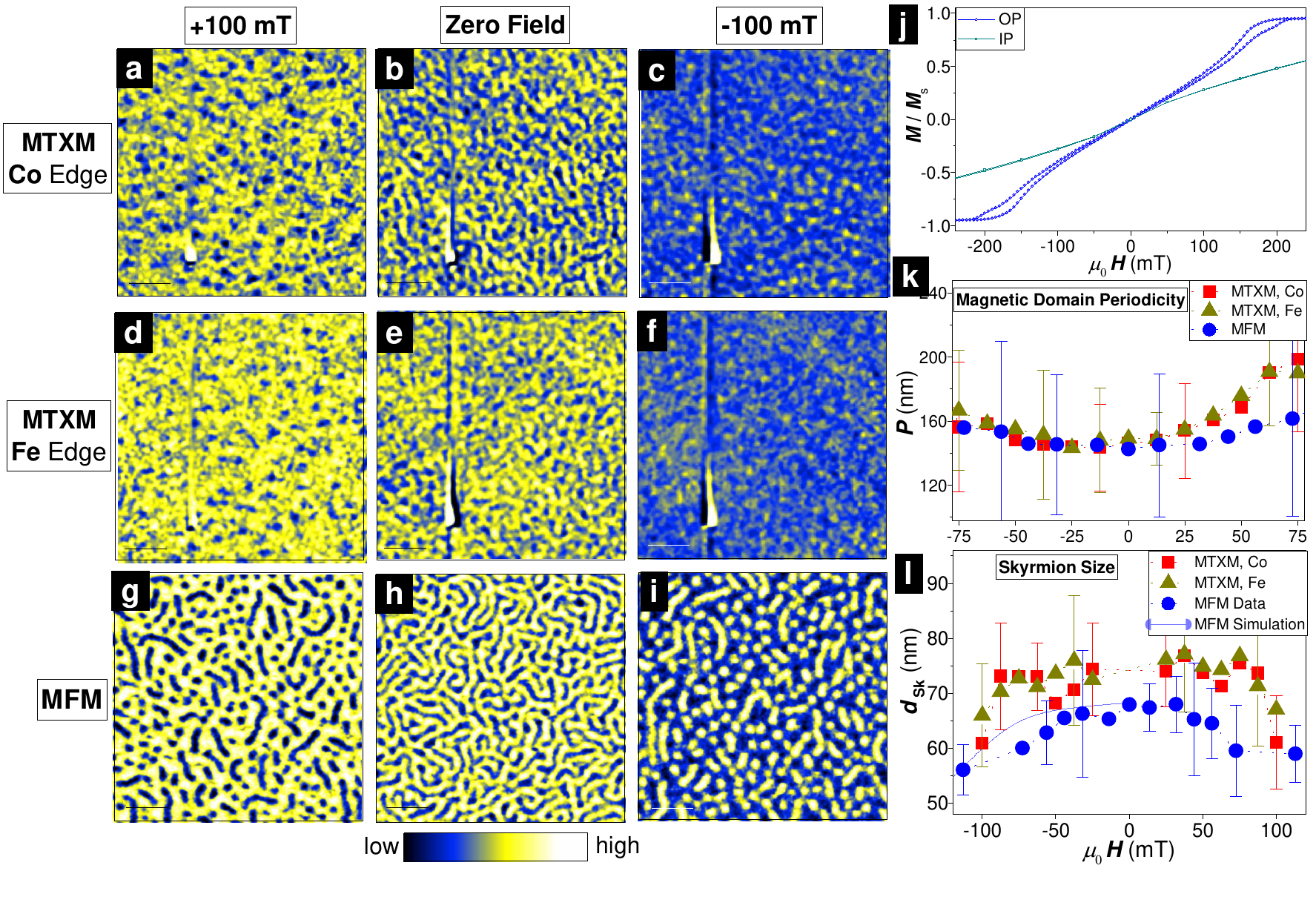}
\par\end{centering}
\caption{\textbf{Magnetic Microscopy of RT Skyrmions.} \textbf{(a-i)} Microscopic
imaging (scale bar: 0.5~\textgreek{m}m) of sample \textbf{Fe(3)/Co(6)}
at RT with MTXM (on Si$_{3}$N$_{4}$ membrane) at the Co L$_{3}$
edge (a-c), MTXM at the Fe L$_{3}$ edge (d-f), over the same sample
region as (a-c), and with MFM (on SiO$_{2}$ substrate, g-i). Images
shown are acquired at $\sim+100$~mT (a, d, g), 0~T (b, e, h) and
$\sim-100$~mT (c, f, i) respectively after saturation at $\sim+250$~mT,
and display similar evolution in magnetic contrast with applied field.
A dead pixel on the MTXM CCD (a-f: bottom left) does not affect our
analysis. \textbf{(j)} Hysteresis loops for out-of plane (OP, blue)
and in-plane (IP, green) magnetization, $M/M_{{\rm S}}$. \textbf{(k-l)}
Comparisons of the field-dependent trends of magnetic domain periodicity,
$P$ (k) and the size of round features (skyrmions) $d_{{\rm Sk}}$
(l), measured in MTXM - Co (red), MTXM - Fe (green) and MFM (blue)
experiments respectively. The size of Néel skyrmions that emerge in
micromagnetic MFM simulations (thick blue line) is overlaid for comparison
(simulation parameters for \textbf{Fe(3)/Co(6)} in \ref{fig:MagTuning}d-f,
details in methods). Representative error bars in (k) indicate the
fit width of $P$, and in (l) the standard deviation of $d_{{\rm Sk}}$
across multiple (minimum 5) skyrmions respectively.\label{fig:MTXM-MFM}}
\end{figure*}

\noindent \textsf{\textbf{}}%

\noindent %
\ref{fig:MTXM-MFM} shows representative images acquired by three
different magnetic microscopy experiments\textbf{ }in varying OP fields
for sample \textbf{Fe(3)/Co(6)}. We begin by examining magnetic transmission
soft X-ray microscopy (MTXM - a technique previously used to study
multilayer skyrmion films\citep{Woo2015}) images acquired using films
grown on Si$_{3}$N$_{4}$ membranes (see Methods). Data recorded
at the Co L$_{3}$ edge (778~eV) (\ref{fig:MTXM-MFM}a-c) show that
as the field is reduced from $+H_{{\rm S}}$, round-shaped sub-100~nm
features with negative (blue) contrast emerge (\ref{fig:MTXM-MFM}a),
which grow to form elongated labyrinthine domains at zero field (\ref{fig:MTXM-MFM}b).
At negative fields, these domains give way to round-shaped features
with positive (yellow) contrast (\ref{fig:MTXM-MFM}c), which shrink
and disappear at $-H_{{\rm S}}$. The sub-100~nm size and field evolution
of these round magnetic features (\ref{fig:MTXM-MFM}l, data) show
striking similarities to the emergence and evolution of Néel-textured
magnetic skyrmions observed in micromagnetic simulations of these
films (\ref{fig:MTXM-MFM}l, line: simulation parameters in \ref{fig:MagTuning}d-f,
details in Methods), and recently reported in similar multilayer films\citep{Moreau-Luchaire2015a,Boulle2016,Woo2015}.
In contrast, we can rule out the presence of magnetic bubbles, which
would be micron-sized and inherently unstable within our films (see
Methods). Therefore, we can establish the identification of these
round magnetic features as Néel skyrmions stabilized by interfacial
DMI.

In contrast to known multilayer skyrmion hosts\citep{Moreau-Luchaire2015a,Boulle2016,Woo2015},
Ir/Fe/Co/Pt stacks include an additional FM layer (Fe). This enables
us to corroborate our observations by performing analogous MTXM experiments
at the Fe L$_{3}$ edge (708~eV, \ref{fig:MTXM-MFM}d-f), over the
same sample region as \ref{fig:MTXM-MFM}a-c. While the magnetic contrast
is diminished (c.f. \ref{fig:MTXM-MFM}a-c), analogous magnetic textures
persist with a comparable field evolution. Importantly, the measured
magnetic domain periodicity and skyrmion size show excellent agreement
across these two experiments (\ref{fig:MTXM-MFM}k-l), confirming
that skyrmionic spin textures persist across the composite (Co/Fe)
FM layer. 

\begin{figure*}
\begin{centering}
\includegraphics[width=4.4in]{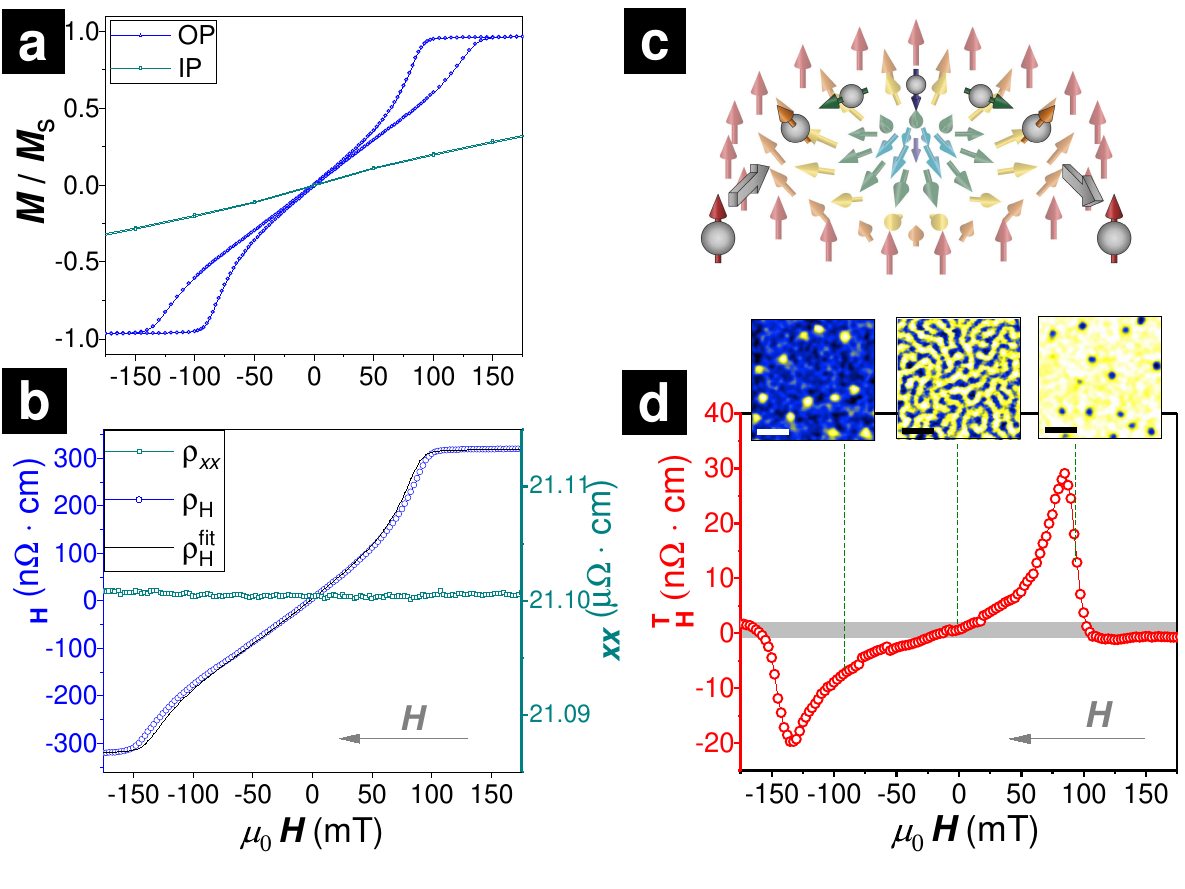}
\par\end{centering}
\caption{\textbf{Topological Hall Effect (THE) from RT Skyrmions. (a)}{\small{}
OP (blue) and IP (green) }magnetization, $M/M_{{\rm S}}$, and (b)
longitudinal ($\rho_{xx}$, green) and Hall ($\rho_{{\rm H}}$, blue)
resistivity as a function of applied field (grey arrow indicates sweep
direction), for sample \textbf{Fe(2)/Co(6)} at RT. The black line
($\rho_{{\rm H}}^{{\rm fit}}$) is a fit to $\rho_{{\rm H}}$, accounting
for conventional and anomalous Hall effects (\ref{eq:THE}\citep{Nagaosa2010}).
\textbf{(c)} Schematic of adiabatic spin rotation of an itinerant
electron (blue sphere, arrow indicates spin orientation) as it traverses
the spin texture of a Néel skyrmion (spectral color scale shows OP
magnetization component). The electron spin undergoes a 2\textgreek{p}
rotation, and the Berry phase accumulated results in the THE. \textbf{(d)
}Residual Hall signal $\rho_{{\rm H}}^{{\rm {\rm T}}}$, ascribed
to THE from Néel skyrmions, obtained from (b) as $\rho_{{\rm H}}^{{\rm T}}(H)=\rho_{{\rm H}}(H)-\rho_{{\rm H}}^{{\rm fit}}(H)$.
Insets show MFM images (scale bar: 0.5 \textgreek{m}m) acquired at
fields corresponding to dashed green lines, indicating the magnetic
structures that generate the THE signal\textbf{.}\label{fig:THE}}
\end{figure*}

To facilitate direct comparison with electrical transport experiments,
we also performed MFM measurements on similar stacks deposited on
Si/SiO$_{2}$ wafers using high spatial resolution tips (diameter
$\sim$30~nm, see Methods). The MFM results (\ref{fig:MTXM-MFM}g-i)
show similar field evolution to MTXM \textendash{} skyrmions at higher
fields (\ref{fig:MTXM-MFM}g, i), and stripes at zero field (\ref{fig:MTXM-MFM}h)
\textendash{} while displaying markedly higher magnetic and spatial
contrast. Despite the difference in substrates (Si/SiO$_{2}$ for
MFM vs. Si$_{3}$N$_{4}$ membranes for MTXM), the MFM domain periodicity
(\ref{fig:MTXM-MFM}k, blue) shows similar magnitude and field evolution
trends as MTXM. 

Finally, we compare in \ref{fig:MTXM-MFM}l the field evolution of
skyrmion size $d_{{\rm Sk}}(H)$ as measured across all three experiments.
The values of $d_{{\rm Sk}}$ reported in \ref{fig:MTXM-MFM}l correspond
to measured values, and thus represent overestimates of the true skyrmion
sizes to varying extents (due to differences in probe sizes and substrates).
The MTXM results (red and green data) show good agreement between
the Co and Fe experiments. Meanwhile, the MFM results (blue data)
are in excellent correspondence with micromagnetic simulations (blue
line), and also agree reasonably well with MTXM \textendash{} albeit
with a systematic offset of $\sim$10~nm. Importantly, the excellent
agreement between the $d_{{\rm Sk}}$ trends across experiments and
simulations firmly establish MFM as a reliable tool for imaging RT
skyrmions. 

\section{Hall Transport Experiments}

\noindent \textsf{\textbf{}}%

\noindent %
To determine the electrical signature of Néel skyrmions, high-resolution
magnetotransport measurements were performed on the films using small,
non-perturbative current densities (as low as $10^{4}$~A/m$^{2}$).
Great care was taken to eliminate any field offsets between the transport
data (\ref{fig:THE}b) and complementary magnetization measurements
(\ref{fig:THE}a, see Methods). The typical RT transport characteristics
for Ir/Fe/Co/Pt stacks (representative sample \textbf{Fe(2)/Co(6)})
are shown in \ref{fig:THE}b, with the longitudinal resistivity ($\rho_{xx}(H)$)
constant to $0.02\%$ within the field range of interest. The Hall
resistivity data, $\rho_{{\rm H}}(H)$ (Fig. 3b, blue), was analyzed
by accounting for contributions from the conventional ($\propto H$)
and anomalous ($\propto M(H)$) Hall effects\citep{Nagaosa2010}:
\begin{equation}
\rho_{{\rm H}}^{{\rm fit}}(H)=R_{0}\cdot H+R_{{\rm S}}\cdot M(H)\label{eq:THE}
\end{equation}
Here, $R_{0}$ is the conventional Hall coefficient, while $R_{{\rm S}}$
represents the cumulative anomalous Hall contribution from skew scattering,
side-jump scattering, and the intrinsic (momentum space) Berry curvature
mechanisms.  After accounting for these contributions, a residual
Hall signal, $\rho_{{\rm H}}^{{\rm T}}(H)=\rho_{{\rm H}}(H)-\rho_{{\rm H}}^{{\rm fit}}(H)$,
is observed with a maximum value of $\sim30$~n\textgreek{W}-cm (\ref{fig:THE}d)
\textendash{} the peak position consistent with an inflection in the
raw $\rho_{{\rm H}}$ data (see \textcolor{blue}{§S3}). Importantly,
the field range corresponding to finite $\rho_{{\rm H}}^{{\rm T}}(H)$
is in good agreement with that over which skyrmions are observed in
MFM (\ref{fig:THE}d insets) and MTXM experiments. We note that in
bulk DMI materials, the observed $\rho_{{\rm H}}^{{\rm T}}(H)$ (5-100~n\textgreek{W}-cm)
has been attributed to Bloch skyrmions\citep{Neubauer2009,Huang2012a,Porter2014}.
Based on the field range, consistency and reproducibility of $\rho_{{\rm H}}^{{\rm T}}(H)$
across samples (§S3), we ascribe its origin to the topological Hall
effect (THE) from Néel skyrmions. 

The THE results from the Berry phase accumulated by itinerant electrons
crossing the 2D skyrmion spin texture (\ref{fig:THE}c). The $\rho_{{\rm H}}^{{\rm T}}(H)$
profile has a characteristic hump shape \textendash{} weak at low
field due to the predominance of 1D stripes (\ref{fig:THE}d, centre
inset), with a peak at higher field due to skyrmion proliferation
(\ref{fig:THE}d, left/right insets), which disappears at saturation\citep{Neubauer2009,Porter2014,Matsuno2016}.
One key contrast with $\rho_{{\rm H}}^{{\rm T}}(H)$ in Bloch skyrmion
materials\citep{Huang2012a,Porter2014} is the observed asymmetry
through the field sweep: in our case $\left|\rho_{{\rm H}}^{{\rm T}}(+H)\right|\neq\left|\rho_{{\rm H}}^{{\rm T}}(-H)\right|$.
Such asymmetry is likely due to hysteretic domain formation driven
by the interfacial anisotropy in our multilayers. Second a direct
comparison between the magnitude of $\rho_{{\rm H}}^{{\rm T}}(H)$
and the observed skyrmion density in MFM images gives an emergent
flux per skyrmion of $6-100\,\phi_{0}$ across our multilayers ($\phi_{0}=h/e$
is the flux quantum, procedural details in §S3). Such a large emergent
flux is in contrast to corresponding reports on Bloch materials ($\mathcal{O}(1)\cdot\phi_{0}$\citep{Neubauer2009,Huang2012a,Porter2014}).
This quantitative discrepancy could be addressed in future by a systematic
comparison between Hall transport and observed magnetic textures.
We note that this is the first reported electrical signature of RT
skyrmions, establishing the utility of Hall transport towards skyrmion
detection in RT devices.

\noindent 
\section{Modulating Magnetic Interactions}

\begin{figure}[h]
\begin{centering}
\includegraphics[width=3.2in]{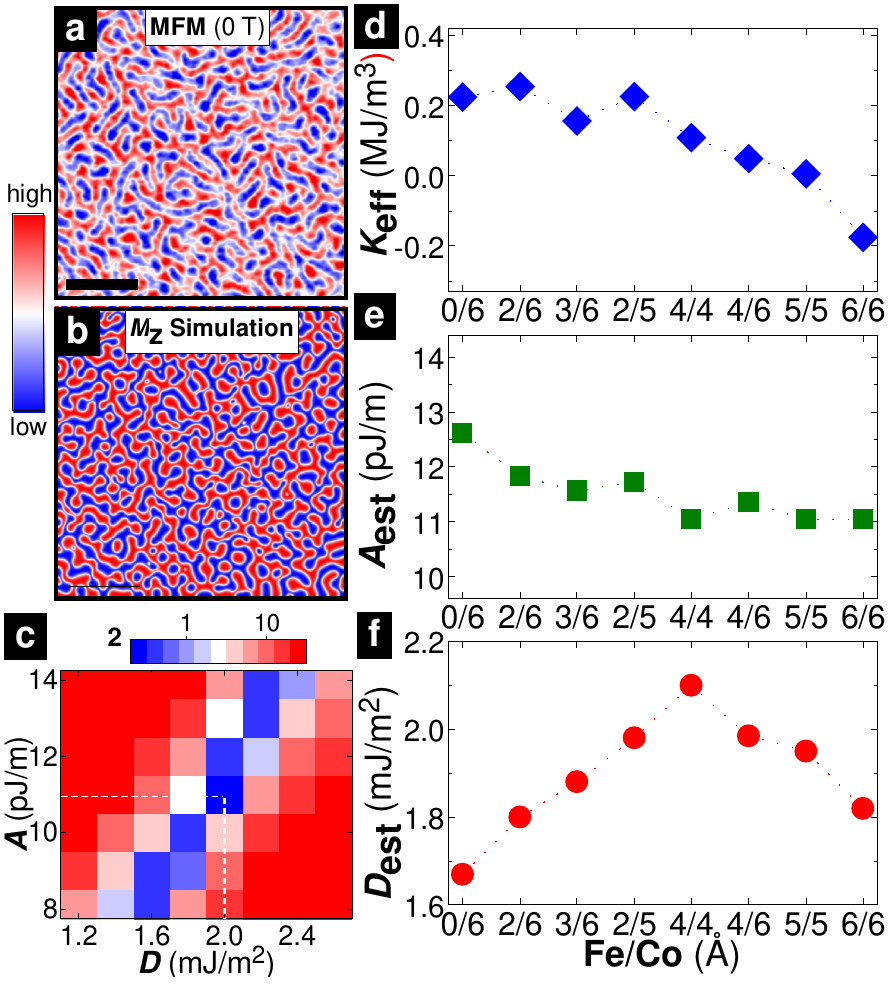}
\par\end{centering}
\caption{\textbf{Modulating Magnetic Interactions with Fe/Co Composition. (a-b)
}MFM image (a) and best-fit micromagnetic simulation ($M_{z}$, b)
(scale bar: 0.5 \textgreek{m}m) of the zero field magnetic contrast
for sample \textbf{Fe(4)/Co(6)}.\textbf{ (c) }A $\chi^{2}$-fit comparison
of the magnetic domain periodicity between MFM data (a) and micromagnetic
simulations, with parameters $A$ and $D$ varied over a range of
likely values. The best fit for \textbf{Fe(4)/Co(6)} corresponds to
$A_{{\rm est}}=11.4$~pJ/m, $D_{{\rm est}}=2.0$~mJ/m$^{2}$ (dashed
white lines, details in methods).\textbf{ (d-f) }The variation of
magnetic interactions $K_{{\rm eff}}$ (d), $A_{{\rm est}}$ (e),
and $D_{{\rm est}}$ (f, obtained using c) across the samples studied
in this work. $D_{{\rm est}}$ is found to be largest for \textbf{Fe(4)/Co(4)},
at $(2.1\pm0.2)$~mJ/m$^{2}$.\label{fig:MagTuning}}
\end{figure}

\noindent %
Having confirmed the presence of RT skyrmions, we examine the effects
of varying \textbf{Fe($x$)/Co($y$)} composition on the magnetic
interactions governing skyrmion properties. First, $K_{{\rm eff}}$
was determined directly from magnetization measurements, and was found
to decrease with increasing Fe/Co ratio and FM thickness (\ref{fig:MagTuning}d).
Next, the exchange stiffness, $A_{{\rm est}}$ was estimated from
DFT calculations (see Methods), and is expected to vary by $\sim15\%$
(11.0 - 12.6~pJ/m) over the Fe/Co compositions studied (\ref{fig:MagTuning}e).
The corresponding DMI, $D_{{\rm est}}$, was determined by comparing
the zero field magnetic domain periodicity in MFM experiments (\ref{fig:MagTuning}a,
for \textbf{Fe(4)/Co(6)}) with corresponding micromagnetic simulations
(\ref{fig:MagTuning}b, see Methods)\citep{Moreau-Luchaire2015a,Woo2015}
using a 2D $\chi^{2}$ fit, with $D$ and $A$ varied over a range
of likely values (\ref{fig:MagTuning}c). Subsequently, the validity
of $D_{{\rm est}}$ and $A_{{\rm est}}$ was established by independently
performing $\chi^{2}$-fits to the field-dependent sssskyrmion size,
$d_{{\rm Sk}}^{{\rm MFM}}(H)$, between experiments and simulations
with varying $D$ and $A$ (see\textcolor{blue}{{} §S4}). Notably, $D_{{\rm est}}$
showed a systematic `dome'-like variation of $\sim30\%$ across samples,
ranging from $1.65-2.1$~mJ/m$^{2}$ (\ref{fig:MagTuning}f) \textendash{}
the maximal value being larger than previous reports for other multilayer
skyrmion hosts\citep{Moreau-Luchaire2015a,Woo2015,Boulle2016}. 

The DMI enhancement observed for Fe($x>0$) relative to\textbf{ Fe(0)/Co(6)}
(Ir/Co/Pt) is consistent with DFT calculations (\ref{fig:Stack-DFT}c).
The inclusion of Fe results in the gradual formation of a Fe/Ir interface
and corresponding suppression of the Co/Ir boundary, leading to increasing
DMI. The DMI reaches its maximal value for complete monolayer coverage
at both Fe/Ir and Co/Pt interfaces (near \textbf{Fe(4)/Co(4)}), and
reduces as the FM layer thickness is further increased due to its
interfacial nature. Notably, the `dome'-like variation in $D_{{\rm est}}$
(\ref{fig:MagTuning}f) does not track the decreasing trend of $K_{{\rm eff}}$
(\ref{fig:MagTuning}d) or that of $A_{{\rm est}}$ (\ref{fig:MagTuning}e).
This indicates that a degree of independent variation of $(D/A)_{{\rm est}}$
and $K_{{\rm eff}}$ can be achieved within our samples. 

\section{Tuning Skyrmion Properties}

\noindent 
\begin{figure*}[t]
\begin{centering}
\includegraphics[width=5.5in]{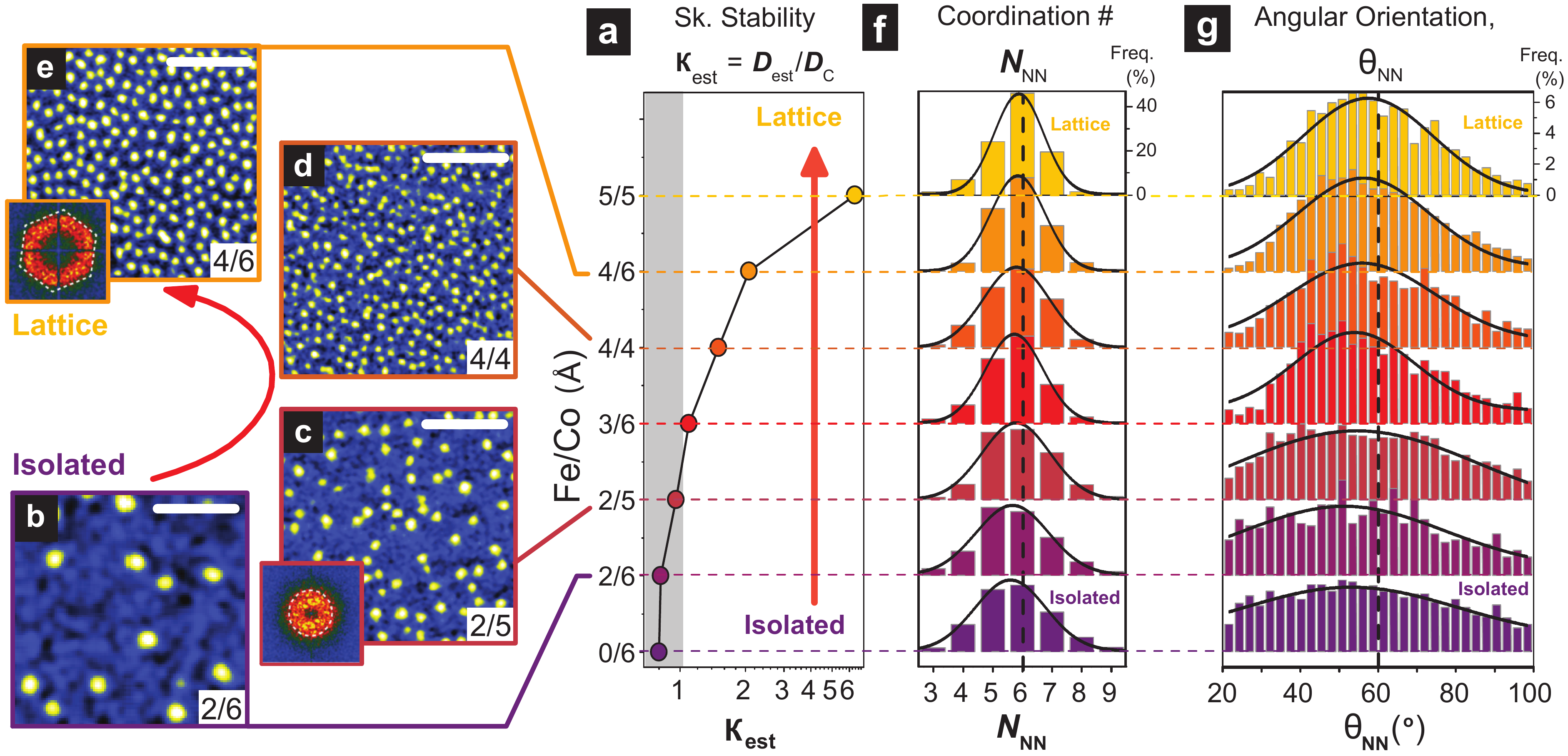}
\par\end{centering}
\caption{\textbf{Tuning Skyrmion Stability with Fe/Co Composition. (a) }Evolution
of the stability parameter $\kappa_{{\rm est}}\equiv D_{{\rm est}}/D_{{\rm c}}$\textbf{
}from $\sim0.8-6.5$ across our samples. \textbf{(b-e) }MFM images
(scale bar: 0.5~\textgreek{m}m) of skyrmion configurations at $H\sim-0.8\,H_{{\rm S}}$
(max. density), for samples \textbf{Fe(2)/Co(6)} (b), \textbf{Fe(2)/Co(5)}
(c), \textbf{Fe(4)/Co(4)} (d), and\textbf{ Fe(4)/Co(6)} (e) respectively.
With increasing $\kappa$, the configuration evolves from sparse,
isolated particles (b) to a dense, disordered lattice (e), with visible
variations in skyrmion size. The short-range hexagonal order in (e)
is evident from the hexagonal Fourier transform (FT, inset), in contrast
to the circular FT for (c) (inset). \textbf{(f-g)} Quantitative analysis
of skyrmion configuration across samples, using Delaunay triangulation
({\small{}details in }\textcolor{blue}{\small{}§S5}{\small{})} to
determine nearest neighbor (NN) statistics (min. sample size: 150
skyrmions): (f) NN coordination number, $N_{{\rm NN}}$, and (g) NN
angular orientation, $\theta_{{\rm NN}}$. With increasing $\kappa$,
the distributions converge around $N_{{\rm NN}}=6$ and $\theta_{{\rm NN}}=60^{\circ}$
(hexagonal lattice). \label{fig:SkConfig}}
\end{figure*}

\noindent %
The observed variation of $D_{{\rm est}}$, $K_{{\rm eff}}$, and
$A_{{\rm est}}$ has direct implications on the properties of skyrmions,
whose stability, size and density evolve visibly across our sample
compositions (\ref{fig:SkConfig}b-e). First, the stability parameter,
$\kappa_{{\rm est}}$ ($\equiv D_{{\rm est}}/D_{{\rm c}}$, \ref{eq:kappa})
varies smoothly from $\sim0.8$ to $\sim6.5$ (\ref{fig:SkConfig}a).
This order of magnitude variation has a dramatic effect on the spatial
configurations of skyrmions near maximal density ($H\sim-0.8\,H_{{\rm S}}$)
(\ref{fig:SkConfig}b-e). The skyrmion configurations can be analyzed
using Delaunay triangulation statistics ({\small{}details in }\textcolor{blue}{\small{}§S5}{\small{})}
for the nearest neighbor (NN) coordination number ($N_{{\rm NN}}$,
\ref{fig:SkConfig}f) and angular orientation ($\theta_{{\rm NN}}$,
\ref{fig:SkConfig}g)\citep{Song2013a}. For $\kappa_{{\rm est}}<1$
(\ref{fig:SkConfig}f-g, bottom plots), the distributions show a large
spread, with a marked deviation from ordered configurations, and correspond
to metastable, isolated skyrmions (circular FT, \ref{fig:SkConfig}c
inset). In contrast, for $\kappa_{{\rm est}}>1$, the distributions
gradually converge around $N_{{\rm NN}}=6$, $\theta_{{\rm NN}}=60^{\circ}$
\textendash{} forming a disordered hexagonal skyrmion lattice (\ref{fig:SkConfig}e
inset). The crossover between isolated and lattice configurations
appears to be smooth, possibly due to the granularity of magnetic
properties and disorder pinning effects in sputtered films. Such a
gradual crossover would likely enable skyrmion stabilization over
a wide range of sizes and densities, while maintaining individual
addressability. 

\noindent 
\begin{figure}[!h]
\begin{centering}
\includegraphics[width=3.2in]{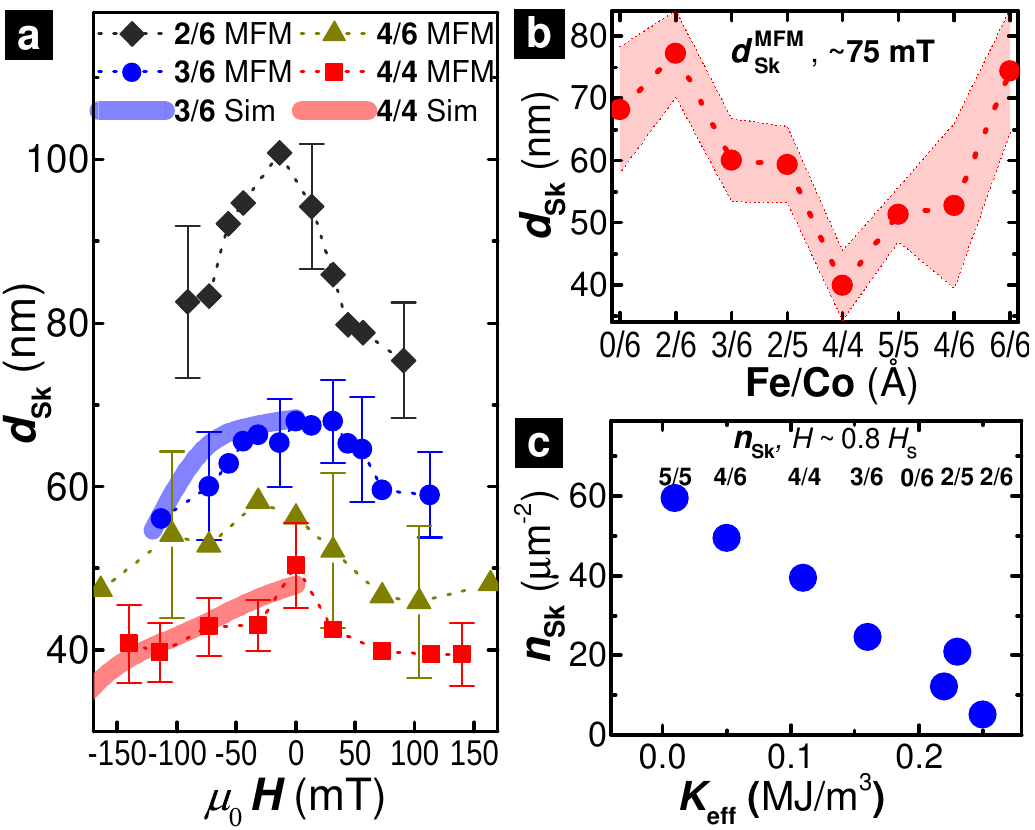}
\par\end{centering}
\caption{\textbf{Tuning Skyrmion Properties with Fe/Co Composition. (a)} Field
dependence of the skyrmion size, $d_{{\rm Sk}}^{{\rm MFM}}$, measured
in MFM images for samples \textbf{Fe(2)/Co(6)} (black), \textbf{Fe(3)/Co(6)}
(blue), \textbf{Fe(4)/Co(6)} (green), and \textbf{Fe(4)/Co(4)} (red).
The MFM results in are compared with corresponding micromagnetic MFM
simulations (parameters from \ref{fig:MagTuning}d-f) for \textbf{Fe(3)/Co(6)}
(blue line) and \textbf{Fe(4)/Co(4)} (red line) respectively.\textbf{
(b)} Variation of $d_{{\rm Sk}}^{{\rm MFM}}$ across samples at $\mu_{0}H\simeq75$~mT
(shaded red region indicates error bars), showing a prominent dip
for \textbf{Fe(4)/Co(4)} and a visible anti-correlation with $D_{{\rm est}}$
(\ref{fig:MagTuning}f). Error bars for (a-b) indicate the standard
deviation of $d_{{\rm Sk}}$ across multiple (minimum 5) skyrmions.
\textbf{(c)} Variation of skyrmion density, $n_{{\rm Sk}}$ with $K_{{\rm eff}}$,
as measured across samples at $H\simeq0.8\,H_{{\rm S}}$. Corresponding
Fe/Co compositions are indicated at the top.\label{fig:SkTuning}}
\end{figure}

\ref{fig:SkTuning}a details the field dependence of the MFM skyrmion
size, $d_{{\rm Sk}}^{{\rm MFM}}$, for four representative samples.\textcolor{blue}{{}
}Notably, skyrmions in \textbf{Fe(4)/Co(4)} ($(D/A)_{{\rm est}}\simeq0.19$:
$d_{{\rm Sk}}^{{\rm MFM}}<$~40~nm) are considerably smaller than
in \textbf{Fe(2)/Co(6)} ($(D/A)_{{\rm est}}\simeq0.14$), \textbf{Fe(3)/Co(6)}
($(D/A)_{{\rm est}}\simeq0.16$), and \textbf{Fe(4)/Co(6)} ($(D/A)_{{\rm est}}\simeq0.17$).
The enhancement of DMI in \textbf{Fe(4)/Co(4)} likely induces a faster
spatial spin rotation, resulting in a large ($>2\times$) reduction
in skyrmion size c.f. \textbf{Fe(2)/Co(6)}. The influence of DMI on
skyrmion size is further evident from the trend in $d_{{\rm Sk}}^{{\rm MFM}}$
across all our samples (\ref{fig:SkTuning}b), which shows a prominent
minimum at \textbf{Fe(4)/Co(4)}, and visible anti-correlation with
the `dome'-shaped variation in DMI (\ref{fig:MagTuning}f). Notably,
the excellent agreement in $d_{{\rm Sk}}(H)$ trends between the MFM
data and corresponding micromagnetic simulations (\ref{fig:SkTuning}a)
underscores the direct relationship between varying magnetic interactions
and skyrmion properties across Fe/Co compositions.\textcolor{blue}{{}
}Finally, we note that the $d_{{\rm Sk}}^{{\rm MFM}}$ values reported
here overestimate the true skyrmion size, due to finite size probe
convolution effects ($\sim$30~nm MFM tips). In fact, RT skyrmions
in \textbf{Fe(4)/Co(4)} could be considerably smaller than any multilayer
skyrmions reported so far\citep{Moreau-Luchaire2015a,Woo2015,Boulle2016}
(see \textcolor{blue}{§S4}). 

While $D/A$ plays a key role in determining the skyrmion size, the
variation in $K_{{\rm eff}}$ is understood to govern the skyrmion
density $n_{{\rm Sk}}$. \ref{fig:SkTuning}c shows the variation
in $n_{{\rm Sk}}$ (for $H\sim0.8\,H_{{\rm S}}$), as a function of
$K_{{\rm eff}}$ across our samples. As $K_{{\rm eff}}$ is reduced
from strongly OP ($\sim$0.3~MJ/m$^{3}$) to near-IP ($\sim$ 0.01~MJ/m$^{3}$),
$n_{{\rm Sk}}$ shows a dramatic increase: from $\sim5$~\textgreek{m}m$^{-2}$
to $\sim60$~\textgreek{m}m$^{-2}$. Skyrmions in low $K_{{\rm eff}}$
samples are stable over a broader range of fields ($>\pm0.2$~T),
due to increased $H_{{\rm S}}$, and are also nucleated close to zero
field with much greater ease. Therefore, varying $K_{{\rm eff}}$
on either side of the optimal DMI composition is a route towards engineering
stacks with the requisite skyrmion density and stability for specific
applications. 

\noindent \textsf{\textbf{}}%

\section{Outlook}

\noindent %
We have synthesized magnetic multilayer stacks hosting RT skyrmions
with tunable properties. By varying the Fe/Co composition, we have
modulated the magnetic interactions governing skyrmion formation,
thereby continuously tuning the thermodynamic stability parameter,
$\kappa_{{\rm est}}$, over an order of magnitude. The resulting skyrmion
configuration, imaged by X-ray microscopy and MFM, evolves from isolated,
metastable particles to a disordered hexagonal lattice. Modulating
$D$, $K$ and $A$ further enables us to vary the skyrmion size (by
$2\times$) and density (by $10\times$). This demonstration of a
platform for tunable sub-50 nm RT skyrmions and their electrical detection
via Hall transport has immediate relevance for device applications. 

The myriad proposals of skyrmion-based memory devices predominantly
build upon either (a) the nucleation or deletion of single skyrmions
in nanostructures\citep{Fert2013,Sampaio2013,Rohart2013}, or (b)
the dynamics of a train of skyrmions in a racetrack configuration\citep{Parkin2008,Tomasello2014,Kang2016}.
The material requirements for (a) would be geared towards the use
of individual, isolated skyrmions in confined geometries. In contrast,
optimal materials for devices based on (b) would correspond to dense
arrays of skyrmions, ideally in an ordered configuration, for high-speed
readout and increased mobility. As demonstrated here, Ir/Fe/Co/Pt
stacks can directly address both these contrasting requirements of
skyrmion properties by simply varying the stack composition. Crucially,
the smooth crossover between isolated and lattice configurations with
composition can simultaneously enable stabilization and individual
addressability for a wide range of skyrmion sizes and densities. We
thus provide a material platform for fast-tracking technological explorations
of skyrmion-based memory devices.

\noindent \bibliographystyle{naturemag}
\bibliography{SkTuning}

\medskip{}

\noindent \textbf{\small{}Acknowledgments}{\small \par}

\noindent {\small{}We acknowledge K. Masgrau, S. He, and B. Satywali
for experimental inputs, W.S. Lew for allowing us to access his instruments,
and P. Fischer, O. Auslaender, and A. Fert for insightful discussions.
We also acknowledge the support of the A{*}STAR Computational Resource
Center (A{*}CRC), Singapore and the National Supercomputing Centre
(NSCC), Singapore for performing computational work. This work was
supported by the Singapore Ministry of Education (MoE), Academic Research
Fund Tier 2 (Ref. No. MOE2014-T2-1-050), the National Research Foundation
(NRF) of Singapore, NRF - Investigatorship (Ref. No.: NRF-NRFI2015-04),
and the A{*}STAR Pharos Fund (Ref. No. 1527400026) of Singapore. M.Y.I.
acknowledges support from Leading Foreign Research Institute Recruitment
Program through the National Research Foundation (NRF) of Korea funded
by the Ministry of Education, Science and Technology (MEST) (2012K1A4A3053565)
and by the DGIST R\&D program of the Ministry of Science, ICT and
future Planning (17-BT-02). The work at ALS was supported by the Director,
Office of Science, Office of Basic Energy Sciences, Scientific User
Facilities Division of the U.S. Department of Energy under Contract
No.DE-AC02-05CH11231.}{\small \par}

\medskip{}

\noindent \textbf{\small{}Author Contributions}{\small \par}

\noindent {\small{}A.S., M.T., F.E., and C.P. designed and initiated
the research. M.R. deposited the films, and characterized them with
A.S. M.Y.I. conducted the MTXM experiments. A.K.C.T. performed the
MFM experiments and analyzed the imaging data with A.S., and P.H.
validated the MFM results. M.R. and A.P.P. performed transport experiments
and analyzed the data with A.S. A.L.G.O. performed micromagnetic simulations.
K.H.K. and C.K.G. carried out the DFT calculations. A.S. and C.P.
coordinated the project and wrote the manuscript. All authors discussed
the results and provided inputs to the manuscript.}{\small \par}

\medskip{}

\noindent \textbf{\small{}Additional Information}{\small \par}

\noindent {\small{}Further details of the results and methods are
provided in Supplementary Information. Correspondence and request
for materials should be addressed to A.S. and C.P. }{\small \par}

\medskip{}

\noindent \clearpage{}

\noindent 
\section{Methods}

\noindent %

\noindent \textsf{\textbf{\small{}Film Deposition}}{\small{}. Multilayer
films consisting of:}{\small \par}

\noindent {\small{}Ta(30)/Pt(100)/{[}Ir(10)/}\textbf{\small{}Fe(0-6)/Co(4-6)}{\small{}/Pt(10){]}$_{20}$/Pt(20)}{\small \par}

\noindent {\small{}(layer thickness in Å in parentheses) were deposited
on thermally oxidized 100~mm Si wafers by DC magnetron sputtering
at RT, using a Chiron\texttrademark{} UHV system (base pressure: $1\times10^{-8}$~Torr)
manufactured by Bestec GmbH. The films were simultaneously deposited
on Si$_{3}$N$_{4}$ membranes (thickness: 200~nm) for MTXM measurements,
and the Ir/Fe/Co/Pt stack was repeated 20 times to enhance the XMCD
contrast. Further deposition and characterization details are in }\textcolor{blue}{\small{}§S1}{\small{}. }{\small \par}

\vspace{0.5ex}

\noindent \textsf{\textbf{\small{}MTXM Experiments}}{\small{} were
performed for several Fe/Co compositions on multilayers deposited
on Si$_{3}$N$_{4}$ membranes. The data were acquired in OP geometry,
with the sample plane normal to the propagation direction of circularly
polarized X-ray beam, using full-field MTXM at the Advanced Light
Source (XM-1 BL 6.1.2). The samples were saturated OP at $+250$~mT,
and MTXM images were acquired through hysteresis loops at the Co L$_{3}$
edge ($\sim$778~eV), and in some cases, at the Fe L$_{3}$ edge
($\sim$708~eV) over the same sample region. The spatial length scales
observed at the Co and Fe edges were calibrated using standard samples,
and the Fe data were corrected for a $\sim7\%$ CCD camera magnification
artifact. }{\small \par}

\vspace{0.5ex}

\noindent \textsf{\textbf{\small{}MFM Experiments}}{\small{} were
performed using a NX-10 AFM/MFM manufactured by Park Systems\texttrademark .
All data were acquired in ambient conditions, with the MFM mounted
on a vibration-isolated platform. The MFM tips used (SSS-MFMR) were
$\sim30$~nm in diameter, with low coercivity ($\sim12$~mT) and
ultra-low magnetic moments ($\sim$80~emu/cm$^{3}$), optimized for
non-perturbative magnetic imaging with high spatial resolution. The
samples were saturated OP using fields up to 0.5~T, and measurements
were performed in OP fields of up to 0.2~T with a typical tip lift
of 30~nm. The contrast of MFM images for $H<0$ has been inverted
for ease of comparison with MTXM images. The zero field results reported
here were consistent with those obtained after AC demagnetization.
Precautions were taken to eliminate tip-induced perturbations, drift,
and other artifacts, and the results obtained were reproduced several
times for consistency. }{\small \par}

\vspace{0.5ex}

\noindent \textsf{\textbf{\small{}Magnetization}}{\small{}, $M(H)$,
of the films was determined using SQUID magnetometry, with a Quantum
Design\texttrademark{} Magnetic Properties Measurement System (MPMS).
The saturation magnetization ($M_{{\rm S}}$) varied over 0.65\textendash 1.1~MA/m
across our films. The effective anisotropy ($K_{{\rm eff}}$) of the
films were determined by acquiring OP and IP hysteresis loops (e.g.
\ref{fig:MTXM-MFM}j), and are detailed in \ref{fig:MagTuning}d. }{\small \par}

\vspace{0.5ex}

\noindent \textsf{\textbf{\small{}Electrical Transport.}}{\small{}
The magnetoresistance and Hall coefficients were measured using a
lock-in technique (excitation frequency: 0-300~Hz), enabling sub-nV
resolution. The measurements were performed using a custom-built variable
temperature insert (VTI) housed in a high-field magnet, complemented
by a Quantum Design\texttrademark{} Physical Properties Measurement
System (PPMS). The data reported here were acquired through a full
hysteresis cycle, with 25~Oe steps within $\pm H_{{\rm S}}$ after
saturation at large fields (+4~T), using small current densities
(as low as $10^{4}$~A/m$^{2}$) so as to not perturb the spin textures.
Importantly, the Hall data were analyzed after carefully accounting
for any magnetic field offsets between magnetization (\ref{fig:THE}a)
and transport (\ref{fig:THE}b) measurements. To this end, repeated
field calibrations were performed on each of the instruments using
reference samples. }{\small \par}

\noindent {\small{}Further procedural details of the analysis and
consistency checks are provided in }\textcolor{blue}{\small{}§S3}{\small{}. }{\small \par}

\vspace{0.5ex}

\noindent \textsf{\textbf{\small{}Density Functional Theory Calculations}}{\small{}.
To estimate the magnetic interactions in our stacks, we performed
first-principles DFT calculations using the technique employed by
Yang }\emph{\small{}et al}{\small{}.\citep{Yang2015a}. The multilayer
stack, Ir{[}3{]}/Fe{[}$b${]}/Co{[}$a${]}/Pt{[}3{]} (number of atomic
layers in braces), were set up in a close-packed (111) orientation
and separated by a vacuum of 10~Å along the OP direction, with the
IP lattice constants set to the calculated bulk Ir value (2.74~Å). }{\small \par}

\noindent {\small{}The DMI was considered only between intralayer
NN atoms to a first approximation, and used to define $d^{{\rm tot}}=\sum_{k}d^{k}$,
i.e. the sum of DMI coefficients, $d^{k}$, for each layer\citep{Yang2015a}.
Subsequently, clockwise and anti-clockwise spin spirals were constructed
across a supercell using the constrained spin method. The energy difference
between these two configurations was computed and scaled by a geometry
and spin-spiral dependent factor to obtain $d^{{\rm tot}}$. This
was used to determine the micromagnetic DMI strength, $D_{{\rm DFT}}$,
across our magnetic films (\ref{fig:Stack-DFT}c). Exchange calculations
performed by first computing $U_{{\rm ex}}$ - the total energy difference
between the spin spiral (averaged over clockwise and anti-clockwise
spirals) and collinear spin configurations. The exchange stiffness
$A_{{\rm DFT}}$ was determined by comparing this exchange energy
density to the micromagnetic exchange free energy.}{\small \par}

\noindent {\small{}Further procedural details for the DFT calculations
and corresponding consistency checks are in }\textcolor{blue}{\small{}§S2}{\small{}.}{\small \par}

\vspace{0.5ex}

\noindent \textsf{\textbf{\small{}Micromagnetic Simulations}}{\small{}
were performed using the mumax$^{3}$ software package\citep{Vansteenkiste2014},
which accounts for interfacial DMI. The multilayers were modelled
with a mesh size of $4\times4$~nm$^{2}$ over a $2\times2$~\textgreek{m}m$^{2}$
area for comparison with films. The $z$ length of the discretization
mesh, $L_{z}$, was changed to match the sample's magnetic layer thickness,
with the non-magnetic spacer approximated to a uniform layer. The
stack configuration used (20 repeats) was consistent with experiments,
and the $M_{{\rm S}}$ and $K_{{\rm eff}}$ values used were obtained
from SQUID measurements. Meanwhile, $A$ and $D$ were varied over
a range of likely values to estimate their magnitude by comparing
the zero field domain periodicity and the field dependence of skyrmion
size. MFM images were simulated for direct comparisons with experiment
using the built-in functionality of mumax$^{3}$. The MFM tip magnetization
was modeled as a magnetic dipole of size 20~nm, with the tip lift
set to 30~nm above the surface. }{\small \par}

\noindent {\small{}For the domain periodicity simulations, the initial
magnetization was randomized, and the relaxed (final) magnetization
configuration ($\vec{M}(x,y,N)$) at zero field was Fourier analyzed
to determine the periodicity. The average periodicity of all 20 magnetic
layers, used for comparison with MFM data, was consistent with that
of the top layer. Finally, the periodicity trends with varying $A$
and $D$ were found to be smooth and monotonic, allowing for regression
analysis to accurately estimate the magnetic parameters for each stack
composition.}{\small \par}

\noindent {\small{}Micromagnetic simulations were also performed with
varying $D$ and $A$ to compare the field evolution of skyrmion size,
$d_{{\rm Sk}}$, with experiments across various sample compositions.
After obtaining the zero field magnetization profile, the magnetization
configuration was relaxed at progressively increasing OP magnetic
fields (steps of 15\textendash 25~mT) until saturation to obtain
a `virgin' magnetization curve. The magnetization configurations obtained
at each field were used to simulate the corresponding MFM images.
The skyrmion size $d_{{\rm Sk}}^{{\rm MFM}}(H)$ was determined by
averaging over Gaussian fits to the simulated MFM profiles of several
skyrmions (typically 5-10) identified from magnetization images. }{\small \par}

\noindent {\small{}Finally, simulations were performed to compare
the observed size and field evolution of skyrmions with the corresponding
phenomenology of bubble systems\citep{Malozemoff1979}. When magnetic
bubbles of comparable sizes (40-100~nm) were artificially introduced
in our films, they were found to be highly unstable \textendash{}
both at zero and finite fields \textendash{} immediately transforming
into chiral domain walls or skyrmions. }{\small \par}

\vspace{0.5ex}

\noindent \textsf{\textbf{\small{}Determining Magnetic Parameters.
}}{\small{}The saturation magnetization, $M_{{\rm S}}$, and effective
anisotropy, $K_{{\rm eff}}$ (\ref{fig:MagTuning}d), were determined
from SQUID magnetometry measurements. The exchange stiffness, $A_{{\rm est}}$
(\ref{fig:MagTuning}e), expected to vary by $\sim15\%$ (11.0 - 12.6~pJ/m)
across samples, was estimated from DFT calculations by linear interpolation
of the results obtained with varying Fe/Co compositions (\ref{fig:Stack-DFT}b).
The corresponding DMI, $D_{{\rm est}}$ (\ref{fig:MagTuning}f), was
determined by performing a $\chi^{2}$-fits to the domain periodicity
at zero magnetic field in MFM experiments with micromagnetic simulations
while varying $D$ and $A$. }{\small \par}

\noindent {\small{}Subsequently, the validity of $D_{{\rm est}}$
and $A_{{\rm est}}$ was established independently by performing $\chi^{2}$-fits
to the skyrmion size, $d_{{\rm Sk}}^{{\rm MFM}}(H)$, between experiments
and simulations for various combinations of $D$ and $A$ across the
films (details in }\textcolor{blue}{\small{}§S4}{\small{}).}{\small \par}

\noindent {\small{}The parameters shown in \ref{fig:MagTuning}d-f
were used together with \ref{eq:kappa} to obtain $\kappa_{{\rm est}}$.
The evolution of skyrmion configuration with $\kappa_{{\rm est}}$
was analyzed using Delaunay triangulation techniques (details in }\textcolor{blue}{\small{}§S5}{\small{}).}{\small \par}

\noindent {\small{}}%
{\small \par}

\vspace{0.5ex}

\noindent \textsf{\textbf{\small{}Magnetic Microscopy and Skyrmion
Properties. }}{\small{}To determine the domain periodicity, the MTXM
and MFM images were analyzed in Fourier space. The position and width
of the FT peak were used to quantify the domain periodicity and the
error bar respectively (\ref{fig:MTXM-MFM}k). Meanwhile, the skyrmion
sizes, $d_{{\rm Sk}}$, were determined from MTXM and MFM measurements
by performing 2D isotropic Gaussian fits to skyrmions identified within
a $\sim6\times6$~\textgreek{m}m$^{2}$ field-of-view ($\sim2\times2$~\textgreek{m}m$^{2}$
for MFM simulations). The error bars for $d_{{\rm Sk}}$ represent
a true spread in the observed size of multiple skyrmions in our films,
consistent with reports in confined geometries\citep{Woo2015}. The
numbers for $d_{{\rm Sk}}$ reported here are raw estimates of skyrmion
width \textendash{} no deconvolution has been performed to account
for the X-ray point spread function or beam profile, or the MFM tip
size.}{\small \par}

\noindent {\small{}}%
{\small \par}

\end{document}